\documentclass[twocolumn,twoside,slac_two]{revtex4}
\usepackage{graphicx}
\usepackage{fancyhdr}
\begin{document}

\title{Flaring Activity from 0836+710 (4C +71.07): What Can We Learn With Limited Multiwavelength Coverage? }

%

\author{A. Akyuz\footnote{on behalf of the {\it Fermi} LAT Collaboration}}
\affiliation{University of Cukurova, Department of Physics, 01330 Adana, Turkey, and NASA Goddard Space Flight Center, Greenbelt, MD 20771, USA}

\author{D. J.Thompson$^{\star}$}
\affiliation{NASA Goddard Space Flight Center, Greenbelt, MD 20771, USA}

\author{D. Donato$^{\star}$}
\affiliation{CRESST, Department of Astronomy, University of Maryland, College Park, MD 20742, USA, and NASA Goddard Space Flight Center, Greenbelt, MD 20771, USA}

\author{L. Fuhrmann\footnote{on behalf of the F-GAMMA Collaboration, http://www.mpifr-bonn.mpg.de/div/vlbi/fgamma/fgamma.html}, K. Sokolovsky}
\affiliation{Max-Planck-Institut f\"ur Radioastronomie, Auf dem H\"ugel 69 53121 Bonn, Germany}

\author{O. Kurtanidze}
\affiliation{Abastumani Observatory, 383762 Abastumani, Republic of Georgia }

\begin{abstract}
After a long period of quiescence in $\gamma$ rays, blazar 0836+710 (4C +71.07) flared in the Spring of 2011. 
 We found only limited multiwavelength coverage of the source.
  An indication of correlated optical/$\gamma$-ray variability is not surprising for a Flat Spectrum Radio Quasar (FSRQ) like this one.  
 Radio observations at high frequencies, however, had seen a flare in 2010, well offset from possible $\gamma$-ray activity.  
The 2011 $\gamma$-ray activity comes during a period of rising radio emission, a pattern that has been seen since the EGRET era. 

\end{abstract}

\maketitle

\thispagestyle{fancy}


\section{Introduction}
The luminous high-redshift (z=2.218) quasar 0836+710 (also known as S5 0836+71 or 4C +71.07)
 is characterized by a flat radio spectrum ($\alpha$=$-$0.33) \cite{Kuhr81}. It hosts a powerful one-sided radio jet emerging from the core 
and extending up to kiloparsec scales \cite{Hummel92}. Very Long Baseline Interferometry (VLBI) images of the source
show a complex motion pattern, with one-sided jet components moving from apparent subluminal to 
superluminal velocities \cite{Otterbein1998}.  Monitoring with MOJAVE has shown apparent speeds up to ($\beta_{app} = 25$, with $\beta = v/c$, $H_{0} = 71~km~s^{-1}~Mpc^{-1}$, $\Omega_m = 0.27$, $\Omega_{\Lambda}=0.73$ \cite{Lister2009}).  
Internal
 structure of the jet in 0836+710  has been investigated at 1.6 and 5 GHz using observations with the VLBI Space Observatory Programme (VSOP) \cite{Lobanov1998}, showing details of the curved jet. 
VLBA observations also suggest a spine-sheath structure for the jet of this source \cite{Asada2010}.
See also the MOJAVE\footnote{http://www.physics.purdue.edu/astro/MOJAVE\\
/sourcepages/0836+710.shtml} and Boston University\footnote{http://www.bu.edu/blazars/VLBAproject.html} summary web pages for this blazar. 
Blazar 0836+710  was also subjected to several X-ray studies and multiwavelength modeling to understand how X-ray emission is produced and the relationship between X-ray emission and other bands \cite{Fang2001, Foschini2006, Gianni2011}.

\section{Gamma-ray Observations}

In the 1990's this blazar was detected and shown to be variable by EGRET on the {\it Compton Gamma Ray Observatory (CGRO)} at  100 MeV $\gamma$-ray energies \cite{Thompsonetal1993}, with the name in the Third EGRET Catalog 
 3EG J0845+7049 \cite{Hartman99}.  Data from other  CGRO instruments, BATSE, OSSE, and COMPTEL at lower $\gamma$-ray energies \cite{Malizia2000, Collmar2006} showed that the peak of the Spectral Energy Distribution (SED) falls in the MeV energy range. A new era in $\gamma$-ray astrophysics began with the launch of the {\it Fermi Gamma-ray Space Telescope (Fermi)} in June, 2008.  This source was not bright enough 
to be included in the {\it Fermi} Large Area Telescope (LAT) Bright Source List  \cite{Abdo2009}; however, it was associated with 1FGLJ0842.2+7054 
in the First LAT Catalog (1FGL, \cite{Abdo1FGL}) and first LAT Active Galactic Nuclei (AGN) Catalog (1LAC, \cite{Abdo1LAC}. Regular $\gamma$-ray monitoring by {\it Fermi}-LAT showed that the source was
 not active until recently.  An exceptional outburst from the source on 2011 April 3 was noted by the 
{\it Fermi}-LAT \cite{Ciprini2011}. Preliminary analysis of observed $\gamma$-ray flare indicated that 0836+710  was in a
 bright state with an average daily flux of F$_{E>100MeV}$$\simeq$ (1.2$\pm$0.3)$\times$10$^{-6}$ photons cm$^{-2}$s$^{-1}$. The reported flux is an order of magnitude greater than the average flux value in the 1FGL catalog.  Its 2FGL name is 2FGL J0841.6+7052 \cite{Nolan2012}.
Its $\gamma$-ray spectrum is steep, with a photon power-law index of $2.95\pm$0.07 in 2FGL. 

\begin{figure*}
\centering
\includegraphics[width=155mm]{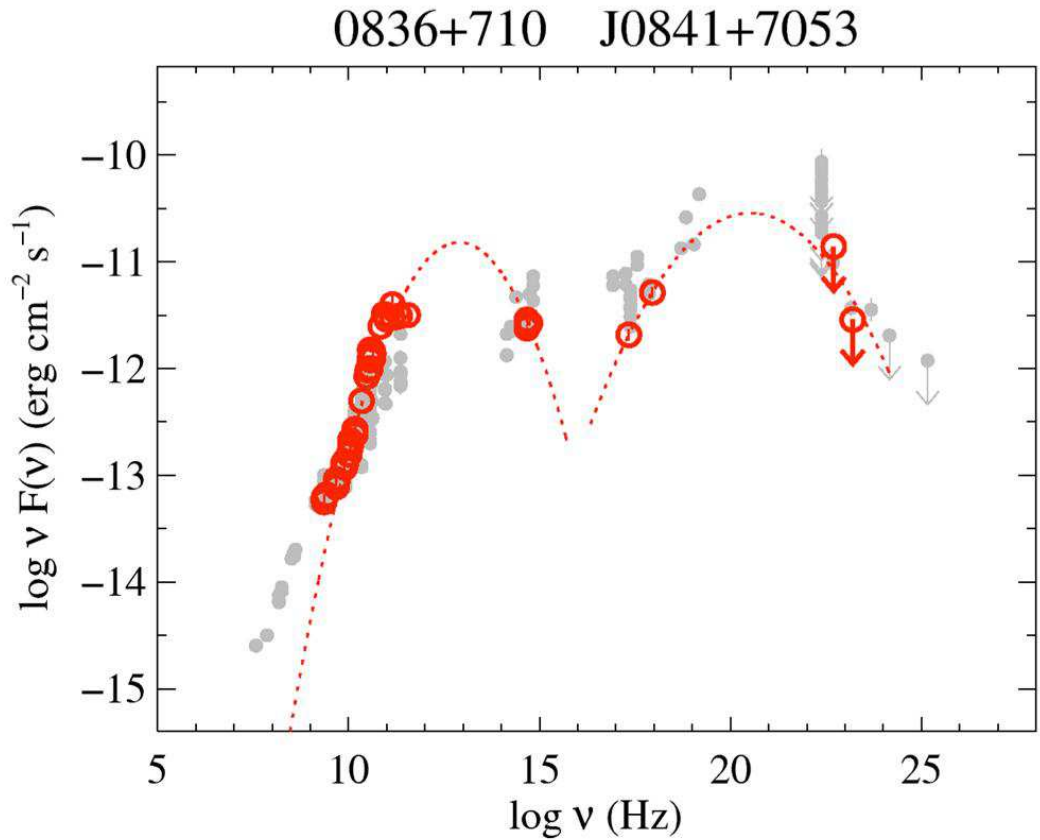}
\caption{The 0836+710  SED shown is from the early {\it Planck} results \cite{Planck Collaboration2011}. 
 During the simultaneous 2010 observations of {\it Planck} and {\it Fermi } (shown in red, along with other simultaneous observations), the $\gamma$-ray flux was not strong enough to produce a good energy spectrum.   
Historical results (gray) show that this blazar has its synchrotron and Compton peaks at relatively low frequencies. The dotted lines are second-order polynomial fits to the two components of the SED.} 
\end{figure*}

\section{Spectral Energy Distribution}

Simultaneous or near-simultaneous SEDs of many Active Galactic Nuclei, including 0836+710, were recently constructed from  the {\it Planck} Early Release Compact Source Catalog (ERCSC) \cite{Planck Collaboration2011}, using a wide range of ground-based and space-based observatories, including {\it Swift} (optical/ultraviolet/X-ray) and {\it Fermi}.  The result for 0836+710 is shown in Figure 1, with simultaneous observations shown in red and historical data in gray.  The {\it Planck} scan for this figure began on 2010 March 17.  During the 2-month interval centered on this scan, the detailed energy spectrum of the blazar was not measurable by {\it Fermi}-LAT (shown as upper limits), because it was still in its quiescent state. The well-known double peak spectra for blazars is seen in the historical data, and the second order polynomial fits to the low-frequency (synchrotron) peak and high-frequency (inverse Compton) peak are shown as dotted lines. 

In the absence of a strong detection in the {\it Fermi }-LAT band during this quiescent period and the lack of broadband simultaneous coverage at times when the $\gamma$-ray emission was detectable, we chose not to attempt a physical model of the SED for this early analysis.  Modeling should be more practical when results from later observations are available (see below).

\section{{\it Fermi}-LAT Analysis}

The {\it Fermi}-LAT data (E$>$100 MeV) considered for this analysis cover the period from 2008 August 4 to 2011 July 4. Only ``diffuse"  LAT events, those likely to be photons, were used. To avoid contamination from the $\gamma$-ray-bright Earth limb, the selection of events with zenith angle $<$105$^{\circ}$ was applied. The  data analysis was performed with the standard analysis tool {\it gtlike}, along with standard Galactic and isotropic diffuse radiation models, all provided with the {\it Fermi}-LAT Science Tools package (v9r23p1)\footnote{Available from the Fermi Science Support Center, http://fermi.gsfc.nasa.gov/ssc/}. The Instrument Response Functions (IRF)
P6$\_$V3$\_$DIFFUSE were used. We restricted the analysis to a region of interest centered on the source and a radius of 10$^{\circ}$.
The source model includes point sources from the 1FGL catalog \cite{Abdo1FGL}, a component for the Galactic diffuse emission (gll\_iem\_v02.fit), and an isotropic component 
(isotropic\_iem\_vo2.txt) that represents the extragalactic diffuse emission as well as residual background, largely from mis-identified cosmic-ray particles.  

Despite the short, bright $\gamma$-ray flare in April, 0836+710 was too faint  in most of the LAT data for detailed studies on short time scales.  We chose, therefore, to concentrate on the long-term behavior.  In the top panel of Figure 2, the source variability is shown by producing a light curve with 4-week time binning and at E$>$100 MeV, starting with the beginning of {\it Fermi} science operations.  Because this is a known source, we calculated flux values for all bins with a Test Statistic (TS) greater than 4, where TS = 2$\Delta$ log(likelihood) between models with and without the source.  As the figure shows, all the bins exceeded this level, indicating that 0836+710 has a continual low-level emission of $\gamma$ radiation.  The $\gamma$-ray light curve shows three broad peaks of emission, centered on MJD 54875 (2009 February), 55360 (2010 June), and 55660 (2011 April). No significant variation in the shape of the $\gamma$-ray energy spectrum was seen during the observations, although the uncertainties are large for these low-level detections. 

\section{Long Term Multiwavelength Light Curve}

Although 0836+710 is monitored regularly by radio telescopes, its coverage at other wavelengths has been fairly sparse in recent years.  Figure 2 summarizes the long-term flux history of the source, anchored in the top panel by the results from {\it Fermi}-LAT and in the bottom two panels by the large F-GAMMA (Fermi-GST AGN Multi-frequency Monitoring Alliance) radio data program \cite{F-Gamma}.  

Contributions to this multiwavelength light curve, in addition to the LAT $\gamma$-ray results, are:

\begin{itemize}
\item Second panel: {\it Swift} X-Ray Telescope observations are taken from the {\it Swift} team automated monitoring program for {\it Fermi} sources\footnote{http://www.swift.psu.edu/monitoring/}, with manual checks on the results near the time of LAT flaring. 
\item Third panel: {\it Swift} Ultraviolet/Optical Telescope observations have been made sporadically in the U, B, and V bands.  We analyzed the public data using the standard tools from the {\it Swift} Center of HEASARC\footnote{http://heasarc.gsfc.nasa.gov/docs/swift/analysis/}. 
\item Fourth panel: R-band optical data come from the blazar monitoring program of the Abastumani Observatory, reduced using their standard software. 
\item Fifth panel: High radio frequency radio data, from the F-GAMMA collaboration, come from the Instituto de Radio Astronom'a Milim\'etrica (IRAM) 30-m telescope (142 and 86 GHz) and the Effelsberg 100-m telescope (42, 32, and 23 GHz).
\item Sixth panel: Lower radio frequency radio data, also from the F-GAMMA collaboration using the 
Effelsberg telescope. 
\end{itemize}

\begin{figure*}
\centering
\includegraphics[width=155mm]{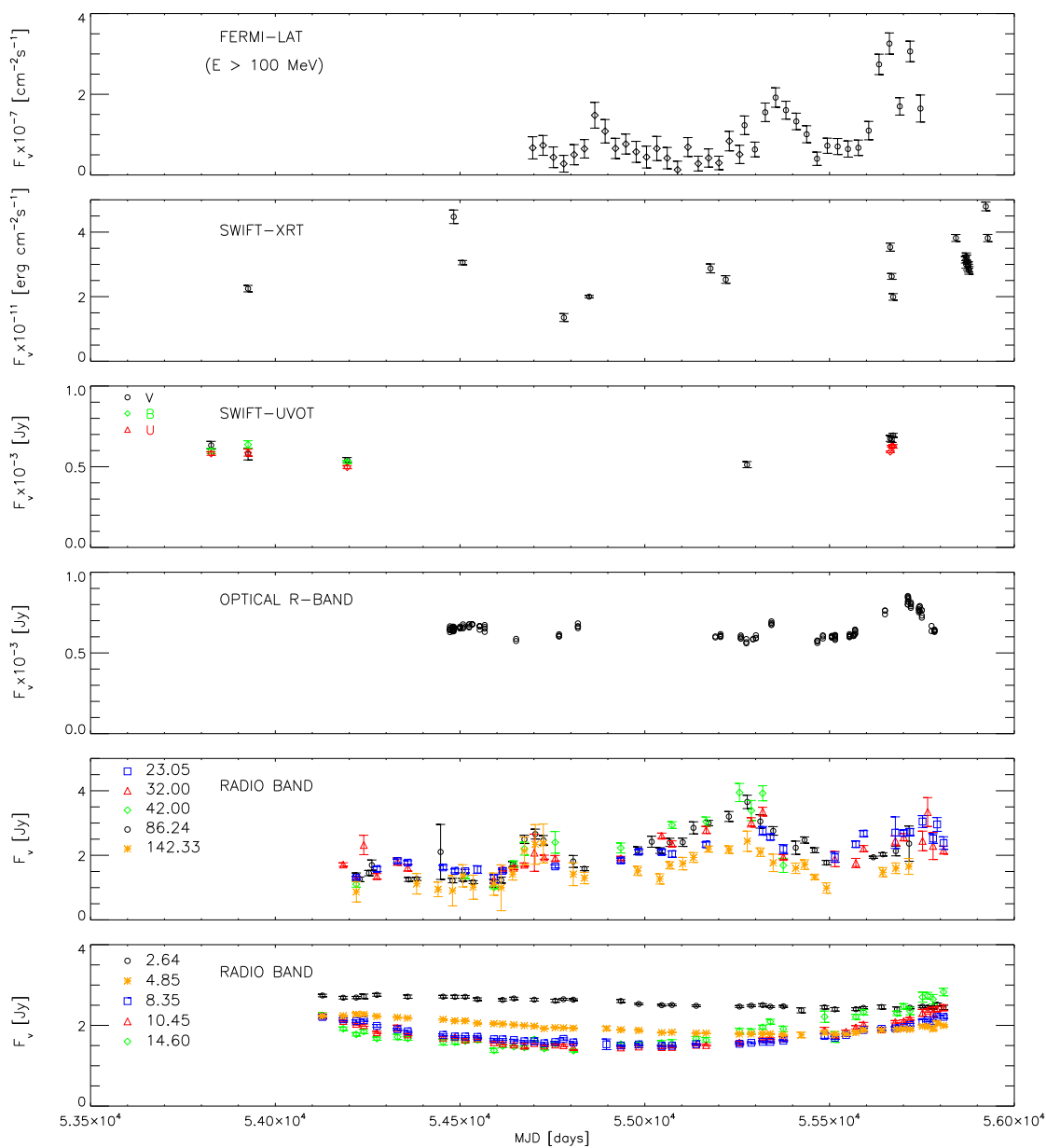}
\caption{Long term multiwavelength light curve of 0836+710.  See text for descriptions of the panels. } 
\end{figure*}

Key features of the long-term light curves: \\

\begin{enumerate}
\item The coverage is sparse except for $\gamma$ rays and radio; therefore
 the emphasis is on long-term rather than short-term correlations. \\ 
 \item The {\it Swift} X-ray Telescope observations 
 show fading X-ray emission following the bright April 2011 $\gamma$-ray flare.  Because the X-ray observations were started after the short $\gamma$-ray flare, a detailed correlation of the time variability in these two bands was not feasible. \\
\item The optical data indicate a correlation 
with the $\gamma$-ray flaring in both 2010 and 2011.  The correlation establishes an identification of the $\gamma$-ray source with 0836+710, but coverage is not 
good enough for a detailed cross-correlation analysis before the 2011 flare. \\ 
\item The high-frequency ($>$ 23 GHz) radio observations also show strong variability with up to a factor of 4 at mm-bands.  Some features:\\
   a) There is a long-term increasing trend starting before the beginning of the LAT light curve and continuing throughout the overlapping observations.\\
   b)  There are also 3 radio flares similar to those seen in $\gamma$ rays, although the cross-band relative timing and correspondence (1:1 correlation) is unclear at the moment.  Quantitative analysis of the amplitudes, shapes, and time offsets should be possible in the near future.  \\
   c)  If the 2010 radio peak ($\sim$MJD 55275) is associated with the $\gamma$-ray peak that same year ($\sim$MJD 55360), then the radio would have to 
lead the $\gamma$-ray emission by a significant amount (months). The observed radio-$\gamma$-ray offset is even longer between the 2008 radio peak ($\sim$MJD 54700) and
 the 2009 $\gamma$-ray maximum ($\sim$MJD 54875).  Detailed correlation studies are in progress. \\
   d)  The stronger $\gamma$-ray flaring activity in 2011 (after MJD 55600) does appear to coincide with a rising flux in the radio, a pattern that has been seen since the EGRET era, e.g. \cite{Valtaoja}. \\
\item The lower-frequency radio bands do not exhibit any significant flaring activity. Except at the lowest (2.64 GHz) radio frequency, all the radio bands show a generally rising trend during most of the {\it Fermi} observations, at least through the middle of 2011.   
\end{enumerate}

On Nov. 1, 2011, 0836+710  flared in $\gamma$ rays to a flux above 100 MeV greater than 3$\times$10$^{-6}$ photons cm$^{-2}$s$^{-1}$, nearly 40 times brighter than the average flux in the 2FGL catalog.  The rise was a factor of 5 in 24 hours or less.  A follow-on paper to these proceedings will incorporate these recent results.

\begin{acknowledgements}

The $Fermi$ LAT Collaboration acknowledges generous ongoing support from a number of agencies and institutes that have supported both the development and the operation of the LAT as well as scientific data analysis. These include the National Aeronautics and Space Administration and the Department of Energy in the United States, the Commissariat \`a l'Energie Atomique and the Centre National de la Recherche Scientifique / Institut National de Physique Nucl\'eaire et de Physique des Particules in France, the Agenzia Spaziale Italiana and the Istituto Nazionale di Fisica Nucleare in Italy, the Ministry of Education, Culture, Sports, Science and Technology (MEXT), High Energy Accelerator Research Organization (KEK) and Japan Aerospace Exploration Agency (JAXA) in Japan, and the K.~A.~Wallenberg Foundation, the Swedish Research Council and the Swedish National Space Board in Sweden.

This research is partly based on observations with the 100-m telescope of the 
MPIfR (Max-Planck-Institut f\"ur Radioastronomie) at Effelsberg. This work has 
made use of observations with the IRAM 30-m telescope. 
\end{acknowledgements}

\end{document}